\documentclass{article}

\usepackage{amsfonts}

\newcommand{\form}[1]{{\tt #1}}

\newcommand{\limplies}{\rightarrow}

\title{Type inference in mathematics}
\author{Jeremy Avigad\footnote{This work has been partially supported by NSF grants DMS-0700174 and DMS-1068829. During the 2009--2010 academic year I spent a sabbatical year working on the verification of the Feit-Thompson theorem with the Mathematical Components group at INRIA-Microsoft Joint Research Centre in Orsay, and I am grateful to Georges Gonthier and the center for support. I am also grateful to Yuri Gurevich, Assia Mahboubi, Enrico Tassi, and an anonymous referee for many helpful comments, corrections, and suggestions.}}

\begin{document}

\maketitle

\begin{abstract}
  In the theory of programming languages, \emph{type inference} is the process of inferring the type of an expression automatically, often making use of information from the context in which the expression appears. Such mechanisms turn out to be extremely useful in the practice of interactive theorem proving, whereby users interact with a computational proof assistant to construct formal axiomatic derivations of mathematical theorems. This article explains some of the mechanisms for type inference used by the \emph{Mathematical Components} project, which is working towards a verification of the Feit-Thompson theorem.
\end{abstract}

\section{Introduction}
\label{intro:section}

Consider the following mathematical assertions:
\begin{itemize}
\item For every $x$ in $\mathbb{R}$, $e^x = \sum_{i = 0}^\infty \frac{x^i}{i!}$.

\item If $G$ and $H$ are groups, $f$ is a homomorphism from $G$ to $H$, and $a$ and $b$ are in $G$, then $f(ab) = f(a) f(b)$.

\item If $F$ is a field of nonzero characteristic $p$, and $a$ and $b$ are in $F$, then
\[
(a + b)^p = \sum_{i=0}^p {p \choose i} a^i b^{p-i} = a^p + b^p.
\]
\end{itemize}
There is nothing unusual about these statements, but, on reflection, one notices that substantial background knowledge and assumptions are needed to parse them correctly. For example, in the first statement, we take it that the index of the summation $i$ ranges over natural numbers, or, equivalently, nonnegative integers. Hence $i!$ is also an integer. Since $x$ is explicitly flagged as a real number, the expression $x^i / i!$ involves division of two different types of objects, taking into account that any integer can be viewed as a real number. In the second statement, $G$ and $H$ are groups, which is to say, each is a set of elements equipped with a group operation and an identity element; so when we write that $a$ and $b$ are in $G$, we really mean that $a$ and $b$ are elements of the underlying set. The notation $ab$ denotes multiplication in $G$, while the notation $f(a) f(b)$ can only be understood in terms of the multiplication in $H$. In the third statement, $p$ is a nonnegative integer (in fact, a prime number, since nonzero field characteristics are prime). But unlike the summation symbol in the first statement, here the summation symbol refers to addition in $F$. In the third statement, ${p \choose i}$ is an integer, while $a^i$ and $b^{p-i}$ are elements of the field. How do we interpret multiplication in \emph{that} case? One way is to notice that there is a canonical map from the integers to any ring with a $0$ and a $1$. Alternatively, any abelian group can be viewed as a $\mathbb Z$-module, which means that it supports scalar multiplication by integers, with all the expected properties; and the additive part of a ring is an abelian group.

Inferences like these are used not only to parse basic mathematical expressions, but also to reason about them correctly. For example, some ``multiplications'' and ``additions'' are commutative, and multiplication often distributes over the corresponding addition. Common manipulations with summations depend on such facts. Understanding mathematics presupposes the ability to keep track of the various domains that objects belong to and variables range over, as well as the relevant operations on those domains and their properties. Our faculties for doing this are so ingrained that we are scarcely aware of the background knowledge we bring to the table when we read an ordinary mathematical text.

The problem is that such background knowledge has to be brought to the foreground when it comes to formalizing mathematics. Broadly speaking, \emph{formal verification} is the practice of using formal methods to verify correctness, such as verifying that a circuit description, an algorithm, or a network or security protocol meets its specification. In this article, I will be concerned, instead, with the verification of mathematical theorems. To be sure, there is no sharp distinction between verifying mathematical statements and verifying claims about hardware and software systems, since the latter are typically expressed in mathematical terms. But ordinary mathematical theorems have a special character, and raise distinct issues and challenges. 

Specifically, I will focus on \emph{interactive theorem proving,} which involves working interactively with a proof assistant to provide enough information for the system to confirm that the theorem in question has, indeed, a formal proof. In fact, many systems actually construct a formal proof object, a complex piece of data that can be verified independently. Systems with substantial mathematical libraries include Coq \cite{bertot:casteran:04} (including the Ssreflect extension \cite{gonthier:mahboubi:10}), HOL \cite{gordon:melham:93}, HOL light \cite{harrison:07c}, Isabelle \cite{nipkow:et:al:02}, and Mizar \cite{grabowski:et:al:10}.  In September 2004, assisted by some students at Carnegie Mellon, I verified a proof of the Hadamard/de la Vall\'ee Poussin prime number theorem \cite{avigad:et:al:07}, using the Isabelle proof assistant. Since then number of nontrival theorems have been formalized, including the four-color theorem \cite{gonthier:08}, the prime number theorem \cite{avigad:et:al:07,harrison:09}, the Jordan curve theorem \cite{hales:07,kornilowicz:07}, G\"odel's first incompleteness theorem \cite{shankar:94,oconnor:05}, Dirichlet's theorem on primes in an arithmetic progression \cite{harrison:09c}, the Cartan fixed-point theorems \cite{ciolli:et:al:11}, and various theorems of measure theory \cite{heller:holzl:11,mhamdi:et:al:11}. There are, moreover, some interesting large-scale verification projects underway. Thomas Hales is heading the \emph{Flyspeck} project \cite{hales:et:al:10}, which aims to verify a number of results in discrete geometry, including the Kepler conjecture. Georges Gonthier is heading the \emph{Mathematical Components} project \cite{garrilot:et:al:09,gonthier:11}, which aims to verify the Feit-Thompson theorem. Fields medalist Vladimir Voevodsky has launched a project to develop ``univalent foundations'' for algebraic topology, providing the basis for formal verification in a theorem prover like Coq.

Checking the details of a mathematical proof is by no means the most interesting or important part of mathematics, and formal verification is not meant to serve as a substitute for mathematical creativity and understanding. But it is generally recognized that the mathematical literature is filled with misstatements, gaps, ambiguities, overlooked cases, omitted hypotheses, and so on, and that the lack of reliability is problematic \cite{nathanson:08}. Moreover, an increasing number of proofs today rely on extensive calculation, and there are currently no standards to ensure that mathematical software is sound. Mathematicians always strive for correctness, and formal verification is simply a technology that is designed to support that goal.

Despite the achievements to date, however, formal verification is still not ``ready for prime time.'' There is a steep learning curve to working with an interactive theorem prover, and verifying even straightforward mathematical results can be frustrating and time consuming. We need better libraries, automated methods, and infrastructure to support verification efforts. This is an exciting time for a young and rapidly evolving field.

In this article, I will focus on one small aspect of formal verification, namely, type inference. In the mathematical setting, the challenge of type inference, roughly speaking, is to keep track of the kinds of objects that appear in a mathematical statement and put that information to good use. What is common to the previous examples is that in each case the relevant information can be inferred from context: 
\begin{itemize}
\item In the expression ``$a$ is in $G$,'' the object of the word ``in'' is expected to be a set.
\item In ``$ab$,'' multiplication takes place in the group that $a$ and $b$ are assumed to be an element of.
\item In ``$x^i / i!$,'' one expects the arguments to be elements of a common structure, for which a division operation is defined.
\end{itemize}
Type inference thus involves not only inferring type information, but also inferring data and facts from type considerations. Of course, type inference is central to the theory of programming languages \cite{pierce:02}, and many of the ideas and methods that have been developed there have been transferred to the mathematical setting. But, as will become clear, mathematical type inference has a distinct flavor. Here I will focus primarily on the approach to type inference used in the Mathematical Components project, which relies on a proof language, Ssreflect, and the Coq theorem prover.

In Section~\ref{inference:section}, I will consider what is desired from a mathematical perspective. In Section~\ref{frameworks:section}, I will discuss some of the underlying axiomatic frameworks, and dependent type theory in particular. In Section~\ref{coq:section}, I will describe some of the mechanisms in Coq that are designed to meet the challenges posed in Section~\ref{inference:section}. In Section~\ref{library:section},  I will describe the way some of these mechanisms are used in the Mathematical Components library, and in Section~\ref{limitations:section}, I will briefly indicate some alternative approaches.

\section{Mathematical type inference}
\label{inference:section}

One hallmark of modern mathematics is the tendency to identify mathematical objects as elements of algebraically characterized structures. Such structures, and classes of such structures, can be related in various ways:
\begin{itemize}
\item Structures in one class may be viewed as elements of a broader one. For example, every abelian group is, more generally, a group, and every group is, more generally, a monoid. Sometimes the inclusions are obtained by taking reducts, which is to say, ignoring parts of the structure. For example, the additive part of a ring is an abelian group, while the multiplicative part is a monoid.
\item A particular structure or a structure in one class can often be embedded in a larger structure. For example, the integers can be embedded in the reals, and every integral domain can be embedded in its field of fractions. 
\item Uniform constructions can be used to build elements of one class of structures from elements of another. For example, the units in any ring form a group, under the associated multiplication; the set of automorphisms of a field (or those fixing some chosen subfield) form a group under composition; any metric space gives rise to a topological space determined by the metric; the field of fractions of any integral domain is a field; and the quotient of a group by a normal subgroup is again a group.
\end{itemize}
What makes this perspective useful is that it allows one to transfer insights and results gained from one domain
to another, and apply background knowledge and expertise uniformly in different settings. The challenge for interactive proof assistants is to reap these benefits.

There are various ways that algebraic methods promote efficiency:
\begin{itemize}
\item They allow us to reuse notation. For example, one may wish to use the symbols $0$ and $+$ with respect to the integers, the reals, and arbitrary rings. 
\item They allow us to reuse constructions. For example, summation $\sum_{i \in I} a_i$ in the integers, reals, and an arbitrary ring can be viewed as instances of the same construction, namely an iteration of the corresponding addition. In fact, various ``big'' operations, including multiplication, logical operations of conjunction and disjunction, lattice operations of meet and join, and so on can be viewed as iterations of an associative operation in an arbitrary monoid.
\item They allow us to reuse facts. Various identities involving big operations can be viewed as instances of general laws that can be instantiated in the different settings. For example, some identities involving summations presuppose that the addition is commutative. Other identities hold in the presence of a multiplication that distributes over addition. We implicitly recognize that such facts hold at various degrees of generality, and instantiate them as appropriate.
\end{itemize}
Any proof assistant that is designed to formalize contemporary mathematical arguments should support these types of reuse.

In the theory of programming languages, type inference allows users to omit information that can be inferred from context. For example, if we write $f(i)$ and $i$ is known to range over the integers, we can infer that $f$ is a function from the integers to some other domain. Various kinds of ``polymorphism'' allow one to reuse symbols and code across different domains. In the context of formally verified mathematics, there are really two types of information that can be inferred:
\begin{itemize}
\item data: for example, the appropriate multiplication in an expression $a \cdot b$, or the appropriate summation operation in an expression $\sum_{i \in A} f (i)$.
\item facts: for example, the fact that $(a \cdot b) \cdot c$ is equal to $a \cdot (b \cdot c)$, when the multiplication in the relevant structure is associative.
\end{itemize}
In the next section, we will see that in certain formulations of logic, these two can be understood as instances of a common phenomena. In other words, inferring a fact can be viewed as inferring a special kind of data, namely ``evidence'' or ``the fact'' that the associated claim is true.

To summarize, in interactive theorem proving, type inference may be invoked when the system parses an expression, but also when the user applies a lemma, or searches for a lemma to apply. The goal of type inference is to allow the user to omit information systematically when such information can be inferred from context. Not only does this save time and energy and reduce tedium, but it also ensures that the expressions we type look like the mathematics we are familiar with, lending support to the claim that our formalizations adequately ``capture'' informal mathematical practice.

\section{Dependent type theory}
\label{frameworks:section}

In order to verify mathematical proofs in a given domain, one has to first choose a formal axiomatic framework that is flexible enough to model arguments in that domain. Experience from the last century has shown that the Zermelo-Fraenkel axioms of set theory provides a remarkably robust foundation for mathematics. Indeed, the Mizar system \cite{grabowski:et:al:10}, which has perhaps the most extensive mathematical library, is based on an extension of $\mathrm{ZF}$ known as Tarski-Groethendieck set theory.

But, in set theory, every object is a set, meaning that the axiomatic framework does not distinguish between numbers, functions, structures, and other objects. For the purposes of type inference, it is often useful to have such distinctions built into the underlying formal system. A number of proof assistants today, including HOL \cite{gordon:melham:93}, HOL light \cite{harrison:07c}, and Isabelle \cite{nipkow:et:al:02}, are based on formulations of higher-order logic like Church's simple type theory \cite{church:40}. One starts with basic types, such as a type {\tt nat} of natural numbers and a type  {\tt bool} of boolean truth values, and adds constructors for forming new types. The most important of these are function types: whenever {\tt A} and {\tt B} are types, so is $\mathtt{A} \rightarrow \mathtt{B}$, intended to denote the type of functions from {\tt A} to {\tt B}. One can also allow, for example, product types $\mathtt{A} \times \mathtt{B}$, denoting the type of ordered pairs from {\tt A} and {\tt B}. Most proof systems have additional mechanisms to support the definition of common mathematical data types and structures, and allow ``polymorphic'' variables ranging over types.

The problem with simple type theory, however, is that it is too simple, since ordinary mathematical structures often depend on parameters. For example, for each $n$, ${\mathbb R}^n$ is a vector space, and for every $n \geq 1$, the integers modulo $n$ form a ring. Thus one may wish to have types
\begin{itemize}
 \item \form{list A n}, denoting sequences of objects of type $A$, with length $n$; and
 \item \form{Zmod n}, denoting the ring of integers modulo $n$.
\end{itemize}
In \emph{dependent type theory}, types can depend on parameters in this way. Notice that such a move tends to blur the distinction between types and terms. For example, in \form{list A n}, the first argument is supposed to denote a type, whereas the second argument is supposed to be a term of type \form{nat}. In some presentations of type theory, this is achieved by having special types, called \emph{universes}, whose terms are also construed as types (see, for example, the presentation of Martin-L\"of type theory in \cite[Section 7.1]{troelstra:van:dalen:88}). Contemporary presentations more often take types to be inhabitants of a third level of syntactic objects, known as ``sorts'' or ``kinds'' (see \cite{barendregt:91}). The specific details need not concern us here; only the fact that terms as well as types can depend on parameters that are again terms or types.

In dependent type theory, the type $\mathtt{A} \to \mathtt{B}$ of functions which take an argument in {\tt A} and return a value in {\tt B} can be generalized to a dependent product $\prod_{\mathtt{x :A}} \mbox{\tt B(x)}$, where {\tt B(x)} is a type that can depend on {\tt x}. Intuitively, elements of this type are functions that map an element {\tt a} of {\tt A} to an element of {\tt B(a)}. When {\tt B} does not depend on {\tt x}, the result is just $\mathtt{A} \rightarrow \mathtt{B}$. Similarly, product types $\mathtt{A} \times \mathtt{B}$ can be generalized to dependent sums $\sum_{\mathtt{x : A}} \mbox{\tt B(x)}$. Intuitively, elements of this type are pairs \form{(a, b)}, where \form{a} is an element of \form{A} and \form{b} is an element of \form{B(a)}. When {\tt B} does not depend on {\tt x}, this is just $\mathtt{A} \times \mathtt{B}$.

In the next section, we will consider one particular theorem prover, Coq. Coq's underlying logic is a dependent type theory known as the \emph{calculus of inductive constructions}, or \emph{CIC} \cite{coquand:paulin:mohring:90}, which extends the original \emph{calculus of constructions} due to Coquand and Huet \cite{coquand:huet:88}. The calculus of inductive constructions has four distinguishing features:
\begin{itemize}
 \item It is a powerful and expressive dependent type theory.
 \item It incorporates the ``propositions as types'' correspondence.
 \item It is constructive, in that every expression in the system has a computational interpretation.
 \item The computational interpretation of terms is used in type checking.
 \item Type checking is decidable.
\end{itemize}
These features are not to everyone's taste, and we will see in Section~\ref{limitations:section} that other proof assistants can reject any or all of them. I will elaborate on each, in turn.

One striking feature of the Calculus of Inductive Constructions is that there are only two basic type-forming operations: dependent products and inductive types. We have already discussed dependent products. Inductive types allow one to define structures that can be characterized as the closure of a set under some basic operations, like the natural numbers, or lists and trees over a type. But, in the CIC, the construction is general enough to include dependent sums, as well as to interpret basic logical notions, like conjunction, disjunction, universal and existential quantification, and equality. In fact, the system has the logical strength of strong systems of set theory \cite{werner:97}.

In order to interpret logical operations in terms of type-theoretic constructions, the CIC relies on what has come to be known as the Curry-Howard ``propositions as types'' correspondence. The point is that logical operations look a lot like operations on datatypes. For example, in propositional logic, from {\tt A} and {\tt B} one can conclude $\mathtt{A} \land \mathtt{B}$. One can read this as saying that given a proof \form{a} of {\tt A} and a proof \form{b} of {\tt B} of $B$ one can ``pair'' them to obtain a proof \form{(a, b)} of $\mathtt{A} \land \mathtt{B}$; or given the ``fact'' \form{a} that \form{A} holds, and the fact \form{b} that \form{B} holds, one obtains the fact \form{(a, b)} that $\mathtt{A} \land \mathtt{B}$ holds. Moreover, from the fact that $\mathtt{A} \land \mathtt{B}$ holds, one can extract the fact that \form{A} holds, and, similarly, \form{B}. If you replace $\mathtt{A} \land \mathtt{B}$ by $\mathtt{A} \times \mathtt{B}$, this is nothing more than a characterization of the product type. In other words, if we posit a new collection {\tt Prop} of types and take the product constructor to map elements {\tt A\ :\ Prop} and {\tt B\ :\ Prop} to {\tt $\mathtt{A} \times \mathtt{B}$\ :\ Prop}, the rules governing products for elements of {\tt Prop} are exactly the desired logical rules for conjunction.

Under this correspondence, implications ${\mathtt{A} \to \mathtt{B}}$ are just instances of function types, and bounded universal quantifiers $\forall {\mathtt{x : A}}. \; \mbox{\tt B(x)}$ are just instances of the dependent product construction. In other words, a proof of $\forall {\mathtt{x : A}}. \; \mbox{\tt B(x)}$ can be viewed as a procedure which, given any object {\tt a : A} returns a proof of {\tt B(a)}. This explains Coq's notation {\tt forall x :\ A, B x} for dependent products. Similarly, the logical construction $\exists {\mathtt{x : A}}. \; \mbox{\tt B(x)}$ is just an instance of the dependent sum. Using inductively defined types, given any type \form{A} one can form \form{$\mathtt{I}_\mathtt{A}$(x,y)~:~Prop} which, intuitively, denotes the proposition that \form{x} is equal to \form{y} as elements of {\tt A}.

One can take the propositions-as-types as expressing a deep insight into the nature and meaning of logical operations \cite{martin:loef:73,tait:86}. But one can just as well view it as a notational convenience which, moreover, allows a proof assistant to treat logical and mathematical operations uniformly. For example, one can take the transitivity of inequality on the natural numbers, \verb\leq_trans\, to be a term of type
\[
 \forall {\mbox{\tt x:nat, y:nat, z:nat,}} \; \mathtt{x} \leq \mathtt{y} \limplies \mathtt{y} \leq \mathtt{z} \limplies \mathtt{x} \leq \mathtt{z}.
\]
This last expression, in turn, it a term of type {\tt Prop}. One can view \verb\leq_trans\ not just as the fact that less-than-or-equal is transitive, but also as a function which, given elements \form{x}, \form{y}, and \form{z} in the natural numbers as well as the facts that $\mathtt{x} \leq \mathtt{y}$ and $\mathtt{y} \leq \mathtt{z}$, return the fact that $\mathtt{x} \leq \mathtt{z}$. Thus, given 
\form{a~:~nat}, \form{b~:~nat}, and \form{c~:~nat}, the term
\verb\leq_trans a b c\ denotes the implication $\mathtt{a} \leq \mathtt{b} \limplies \mathtt{b} \leq \mathtt{c} \limplies \mathtt{a} \leq \mathtt{c}$. Moreover, we can express that \form{H} is the fact that $\mathtt{a} \leq \mathtt{b}$ by writing \form{H~:~$\mathtt{a} \leq \mathtt{b}$}, in which case \verb\leq_trans a b c H\ denotes the implication $\mathtt{b} \leq \mathtt{c} \limplies \mathtt{a} \leq \mathtt{c}$.

The propositions-as-types correspondence is particularly popular as a foundation for constructive mathematics, where assertions are expected to have direct computational significance. Every term in Coq can be viewed as a computational object, subject to evaluation. For example, if $\pi_0$ and $\pi_1$ denote the two projections from a product type $\mathtt{A} \times \mathtt{B}$, the a term $\pi_0 \mbox{\tt (a, b)}$ can be ``reduced'' or ``evaluated'' to \form{a}. In fact, every term in Coq can, at least in principle, be reduced to a canonical normal form. In particular, if \form{t} is a closed term of type {\tt nat}, then \form{t} reduces to a numeral. Coq, moreover, makes use of this computational interpretation when checking types. For example, If {\tt C(x)} is a type that depends on a value {\tt x} of type {\tt A}, the system can recognize that {\tt C($\pi_0 \mbox{\tt (a, b)}$)} is the same type as {\tt C(a)}. 

The decidability of type checking amounts to the fact that given a term, {\tt t}, and a type, {\tt T}, the type-checker can, deterministically, decide whether or not {\tt t} has type {\tt T}. This is clearly a useful property to have, though we will see, in Section~\ref{limitations:section}, that it imposes strong restrictions. Under the propositions-as-types correspondence, the decidability of type checking takes on additional significance. Suppose \form{P} is a term of type {\tt Prop}, expressing, for example, Fermat's last theorem. Then a term {\tt t} of type {\tt P} is a proof that {\tt P} is true. Proving Fermat's last theorem thus amounts to constructing a term of type {\tt P}, and the decidability of type checking implies that such a term can be recognized, algorithmically, as a valid proof.

\section{Type inference in Coq}
\label{coq:section}

Now that we have a sense of Coq's axiomatic framework, let us explore some of the mechanisms the system offers to address the challenges raised in Section~\ref{inference:section}. Generally speaking, type inference is triggered when the system is called on to determine the type of a term, or to check that a term has an appropriate type, when some information has been left implicit. But because dependent types depend on the values of their parameters, inferring a type can entail inferring such values. Recall that in Section~\ref{inference:section} we distinguished between two types of information that can be inferred, namely, data and facts. With the propositions-as-types correspondence in place, inferring a fact---such as the fact that multiplication is associative---is a matter of inferring a value of a type {\tt P}, which is in turn of type {\tt Prop}, where {\tt P} expresses the expected associativity property.

We will consider three principal mechanisms. \emph{Implicit arguments} allow users to systematically leave information out of an expression when this information can be inferred from context. \emph{Coercions} allow users to cast, implicitly, objects of one type to objects of another. Finally, \emph{canonical structures} let the user register certain objects as components of a larger structure, providing useful information to the type inference process.\footnote{For more detail than is provided below, see Coq's online reference manual. All three mechanisms were initially introduced to Coq by Amokrane Sa\"{\i}bi \cite{huet:saibi:00,saibi:97,saibi:99}, who credits the idea of using implicit arguments in the theorem proving context to Peter Aczel. Implicit arguments were further extended by Hugo Herbelin and Matthieu Sozeau. Canonical structures received little attention until they were revived and used aggressively by Gonthier; see, for example, \cite{garrilot:et:al:09}.} 

It will be helpful to illustrate these with a running example. The following definition declares a new type, {\tt group}:
\begin{verbatim}
  Record group : Type := Group 
  {
    carrier : Type;
    mulg : carrier -> carrier -> carrier;
    oneg : carrier;
    invg : carrier -> carrier;
    mulgA : forall x y z : carrier, 
      mulg x (mulg y z) = mulg (mulg x y) z;
    ...
  }.
\end{verbatim}
According to this type declaration, \form{group} is a record type, consisting of a carrier, a multiplication, an identity, and an inverse. These are assumed to satisfy the requisite axioms, such as the associativity of multiplication. If \form{G} has type \form{group}, that is, \form{G : group}, then the components of \form{G} are \form{carrier G}, \form{mulg G}, \form{oneg G}, and so on. Conversely, given elements \verb\my_carrier\, \verb\my_mul\, \verb\my_one\ and so on of the right type, the term \verb\Group my_carrier my_mul my_one ...\ denotes the corresponding group.

Notice that we are relying on dependent type theory here. The type \form{group} is a classic example of a dependent sum, since, for example, the type of the second component, \form{carrier -> carrier -> carrier}, depends on the value \form{carrier} of the first component. The arguments of the corresponding projections bear the associated dependences. For example, the term \form{mulg}, which picks out the the second component, has type \form{forall G : group, carrier G -> carrier G -> carrier G}, a dependent product. 

Notice also that the proposition-as-types correspondence is being put to good use. For example, the type of the fifth component, \form{mulgA}, is the proposition that \form{mulg} is associative. Assuming \form{G\ :\ group}, the term \form{mulgA G} has type
\begin{verbatim}
  forall x y z : carrier G, 
    mulg G x (mulg G y z) = mulg G (mulg G x y) z 
\end{verbatim}
which is itself a term of type {\tt Prop}. Thus \form{mulgA G} denotes the fact that multiplication in \form{mulg G} is associative, a fact that can be applied to elements of the carrier of \form{G} just as in the example of \verb\leq_trans\ above. In this way, the propositions-as-types correspondence provides a natural and convenient way to think of the group structure as including not only the relevant data---the carrier of the group and group operations---but also the relevant properties.

In a context where we have \form{G\ :\ group}, \form{g\ :\ carrier G}, and \form{h\ :\ carrier G}, the term \verb=mulg G g h= represents the product of {\tt g} and {\tt h} under the multiplication operation of {\tt G}. The   implicit arguments mechanism in Coq allows us to write \verb\mulg _ g h\, replacing the first argument by an underscore. Doing so means that we expect the type inference algorithm to infer the value of that argument from context, by finding a solution to the constraints imposed by the fact that the resulting term should be well typed. The algorithm proceeds by instantiating the first element with a variable, \form{?}. The term \form{mulg ?} then has type \form{carrier\ ?\ -> carrier\ ?\ ->\ carrier\ ?}. Since this term is applied to \form {g\ :\ carrier G}, to get the types to work out the system has to solve a simple unification problem, namely, instantiating \form{?} to unify \form{carrier\ ?} with \form{carrier G}. Thus \form{?} is instantiated to \form{G}, and the algorithm has inferred the relevant parameter. With this in mind, one can introduce a new notation:
\begin{verbatim}
  Notation "g * h" := (mulg _ g h).
\end{verbatim}
This enables one to write \form{g * h} for multiplication in any group, allowing the group in question to be inferred from the type of \form{g}.

In this example, the implicit argument mechanism was used to infer a parameter in the application of a function, \form{mulg}. But the mechanism can be used just as well to infer parameters during the application of a lemma. For example, recall the transitivity lemma \verb\leq_trans\ from the last section. This takes five arguments---three natural numbers, {\tt x}, {\tt y}, {\tt z}, and the facts $\mathtt{x} \leq \mathtt{y}$ and $\mathtt{y} \leq \mathtt{z}$---and returns the fact $\mathtt{x} \leq \mathtt{z}$. Suppose we declare the first three arguments to be implicit. Then given \form{H1\ :\ $\mathtt{a} \leq \mathtt{b}$} and \form{H2\ :\ $\mathtt{b} \leq \mathtt{c}$}, the term \verb\leq_trans H1 H2\ has type $\mathtt{a} \leq \mathtt{c}$. Moreover, when we are building a proof interactively in Coq, if we apply \verb\leq_trans H1\ to a subgoal $\mathtt{a} \leq \mathtt{c}$, type inference similarly infers the missing arguments and leaves the us with the goal $\mathtt{b} \leq \mathtt{c}$. 

Coercions are commonly used in programming languages, for example, when adding a real and an integer triggers the coercion of the integer to a real. In the context of mathematical theorem proving, coercions have other uses as well. In our running example, one would ordinarily write \form{g\ :\ carrier G} to specify that \form{g} is an element of the carrier of \form{G}. Writing \form{g\ :\ G} instead yields an error, because the system expects something of type {\tt Type} on the right side of the colon, and {\tt G} has type {\tt group}. But declaring
\begin{verbatim}
  Coercion carrier : group >-> Type.
\end{verbatim}
informs Coq that the function \form{carrier} can always be used to coerce a group to a type. If one then enters \form{g\ :\ G}, the algorithm finds itself facing a group on the right side of the colon but expecting a type, and readily inserts the coercion.

The last feature that we will discuss, canonical structures, provides an inverse to coercion, of sorts. In the example above, we used the \form{carrier} function to coerce a record structure to one of its projections. Canonical structures makes it possible for the type inference algorithm to pass in the other direction, and recognize a particular object as the projection of a larger structure. To illustrate, suppose we define
\begin{verbatim}
  IntGroup := Group int addi zeroi negi addiA ...
\end{verbatim}
thereby declaring the integers with addition to be an instance of a group. Somewhat perversely, this will allow us to write \form{mulg IntGroup i j} instead of \form{i + j}, when we have \form{i j\ :\ int}. Less perversely, this will allow us to apply general theorems about groups to this particular instance. But what happens now when we write \form{i * j}? This expression is shorthand for \verb=mulg _ i j=. After instantiating the first argument to a variable, {\tt ?}, the type inference algorithm is faced with the unification problem \form{carrier ?\ = int}, and gets stuck. Declaring
\begin{verbatim}
  Canonical Structure IntGroup.
\end{verbatim}
registers the fact \form{carrier IntGroup = int} with the system for use in type inference. One can view this as a ``hint'' to the unification process \cite{asperti:et:al:09}. Now when the type inference algorithm gets stuck as above, it can appeal to a table of such hints, and use the relevant one to recognize that the integers can be viewed as the carrier of the \form{IntGroup} structure. The algorithm then replaces \form{int} by \form{carrier IntGroup} and solves the unification problem.

The mechanisms just described are not exceedingly complicated, but we will see in the next section that they are remarkably robust with respect to the challenges posed in Section~\ref{inference:section}. Canonical structures can, moreover, be used in clever ways to trick the type inference algorithm into carrying out various kinds of useful automation \cite{gonthier:et:al:11}. 

To summarize, type checking is triggered when the user enters an expression or applies a lemma, possibly leaving some arguments and facts implicit. Coq's type inference engine has four resources at its disposal to fill in the remaining information:
\begin{enumerate}
\item \emph{unification} can be used to infer implicit arguments;
\item \emph{coercions} can be inserted to resolve a type mismatch; 
\item the unification algorithm can refer to a database of \emph{unification hints} to solve unification problems involving a projection of a \emph{canonical structure}; and
\item when all else fails, the algorithm can simplify terms or unfold definitions according to the CIC's computational interpretation of terms, and then retry the previous steps.
\end{enumerate}
Generally speaking, implicit arguments can trigger arbitrary instances of higher-order unification, which is known to be undecidable \cite{dowek:01}. So, at best, type inference can only aim to search a reasonable fragment of the space of possible instantiations for an implicit argument. And even within decidable fragments, unpacking definitions and unfolding terms can easily lead to combinatorial explosion. Nonetheless, Coq's type inference algorithm consists, essentially, of iterating the steps above, relying on heuristics to limit the possibilities in the fourth step.

\section{The mathematical components library}
\label{library:section}

This section provides a brief indication of some of the ways that the mechanisms for type inference discussed in Section~\ref{coq:section} have been used towards Gonthier's formalization of the Feit-Thompson theorem \cite{feit:thompson:63}, which asserts that finite groups of odd order are solvable. These examples only scratch the surface; for more detail, see \cite{bertot:et:al:08,garrilot:et:al:09,gonthier:08,gonthier:mahboubi:10,gonthier:et:al:07}.

Recall that Coq's logic is constructive. In contrast, many principles and methods that are commonly used in contemporary mathematics are not constructively valid. For example, constructively, one cannot assume the law of the excluded middle, or prove the existence of an $x$ satisfying a property $P$ by assuming there is no such $x$ and deriving a contradiction. Extensionality fails: one cannot, in general, prove that two functions $f$ and $g$ from $A$ to $B$ are equal by proving that $f(x) = g(x)$ for every $x$. Choice fails as well: even if one has proved that for every $x$ in $A$ there is a $y$ in $B$ such that some property holds, one cannot assume that there is a function $f$ that picks out such a $y$ for every $x$. 

On the other hand, these properties generally hold in \emph{finite} domains. Since the Feit-Thompson theorem is an extended exploration of properties of finite groups, one would like to take advantage of these features when they are available. Thus, in the Ssreflect library, there are general structures for types with a decidable equality relation (that is, ones where the relation can be computed by a function returning a boolean value of ``true'' or ``false,'' ensuring that it satisfies the law of the excluded middle); finite structures; and structures that can be equipped with choice functions. For example, one can define a structure for types with decidable equality as follows:
\begin{verbatim}
  Record eqType : Type := EqType
  {
    carrier : Type;
    rel : carrier -> carrier -> bool;
    ax : forall x y, (x = y) = (rel x y)
  }.
\end{verbatim}
In the last line of the record, the expression {\tt rel x y} of type {\tt bool} is coerced to the proposition that the value of this expression is equal to {\tt true}. In other words, {\tt ax} is the proposition that {\tt rel x y} holds if and only if {\tt x = y}. Declaring \form{carrier} to be a coercion allows one to write \form{x\ :\ T} whenever we have \form{T\ :\ eqType}. Implicit arguments allow one to use the notation \form{x == y} in place of {\tt rel T x y} whenever \form{x} and \form{y} are elements of the carrier of such a \form{T}. Finally, canonical structures allow one to associate the relevant boolean equality relation with the natural numbers, so that one can write \form{x == y} when we have \form{x y\ :\ nat}, as well. (This is a slight simplification of the implementation in the Ssreflect library \cite{garrilot:et:al:09}.)

Section~\ref{inference:section} noted that ``big operations'' such as $\sum$, $\prod$, $\bigcap$, $\bigcup$, $\bigwedge$, $\bigvee$ can all be viewed as instances of iterations of an associative binary operation. But such operations come in many different flavors: one can sum over a list, a numeric range, or a finite set, and these summations will satisfy different properties depending on whether the underlying structure is a semigroup, an abelian semigroup, or a ring. Ssreflect comes with an overarching ``bigop'' library, and once again type inference plays a key role in making it work \cite{bertot:et:al:08}.

Type inference is also used to manage algebraic class inclusions (between rings, commutative rings, fields, and son on) and algebraic constructions: for example, the set of $n$ by $n$ matrices over a ring forms a ring when $n > 0$, and the set of polynomials over a commutative ring again forms a commutative ring. Type inference ensures that the relevant algebraic facts are readily available, and allows a uniform use of notation \cite{garrilot:et:al:09,gonthier:11a}. Definitions in the Ssreflect library have been carefully chosen so that if \form{G} and \form{H} are groups of the same type (more precisely, subgroups of some ambient group type), then the quotient notation \form{G\ /\ H} makes sense; but when \form{H} is in fact a normal subgroup of \form{G}, as in the usual construction of a quotient group, \form{G / H} is a group with all the expected properties \cite{gonthier:et:al:07}. For another example, when a group \form{G} happens to be abelian, it is often treated as a $\mathbb{Z}$-module and written additively. So, for example, one can write \form{g *+ n} for scalar multiplication of {\tt g} by {\tt n} whenver {\tt g} is an element of the group and {\tt n} is a natural number. Type inference is used to mediate between these two ``views'' of an abelian group.

Type inference also helps with mundane mathematical conventions. For example, Section~\ref{inference:section} noted the conflation of groups with sets. If \form{G} and \form{H} are subgroups of an ambient finite group, and {A} is a subset of that group, then $\mathtt{G} \cap \mathtt{H}$ and \verb\C_G(A)\ (the centralizer of \form{A} in \form{G}) are both groups. But they are also just sets with the ambient group operation; an element \form{x} is in $\mathtt{G} \cap \mathtt{H}$ if and only if it is in \form{G} and \form{H}, and \form{x} is in \verb\C_G(A)\ if and only if \form{x} is in \form{G} and commutes with every element of \form{A}. Type inference mediates between these two views of a construction---that is, of yielding both a group and a set---allowing one to apply lemmas involving groups in some instances and lemmas involving sets in others. For another example, a homomorphism between groups $G$ and $H$ is a function between $G$ and $H$ equipped (using a record type) with additional properties. Coercion allows one to use ordinary function notation with morphisms, such as \form{f x} and $\mathtt{f} \circ \mathtt{g}$. In the other direction, canonical structures automatically infer the fact that $\mathtt{f} \circ \mathtt{g}$ is a homomorphism when \form{f} and \form{g} are, giving $\mathtt{f} \circ \mathtt{g}$ a similarly dual status as function and morphism.

Canonical structures can even be used to make sense of mildly abusive mathematical notation. For example, if $U$ and $W$ are subspaces of a vector space $V$, it is common to write $U + W$ for set $\{ u + w \; | \; u \in U, w \in W \}$. Mathematicians will often say ``$U + W$ is a direct sum'' when $U$ and $W$ have trivial intersection, ignoring the fact that this is a property of the pair $(U, W)$ which is impossible to read off from the $U + W$ alone. Gonthier has shown, however, that canonical structures provide a convenient way of supporting this abuse of language \cite{gonthier:11a}.

\section{Limitations and other approaches}
\label{limitations:section}

The mechanisms supporting type inference that were described in Section~\ref{coq:section} are not the only ones available in Coq. In particular, Coq now has a ``type class'' mechanism \cite{sozeau:oury:08}. Type classes and canonical structures serve similar purposes, but whereas canonical structures are handled within the type inference loop described at the end of Section~\ref{coq:section}, the type class mechanism collects constraints that are passed to a separate inference engine at the end of the process. Spitters and van der Weegen \cite{spitters:van:der:weegen:11} have experimented with type classes in the context of mathematical type inference, with positive results. 

But one may wish to stray even further from Coq's mindset. Recall some of the key features of that proof assistant:
\begin{itemize}
 \item An elaborate type theory is built in to the underlying axiomatic framework.
 \item Using the propositions-as-types correspondence, data and facts are handled in the same way, so theorems can be applied to arguments and hypotheses just as functions are applied to arguments.
 \item The underlying logic is constructive, and every term has computational significance.
 \item Type checking makes use of the computational interpretation of terms.
 \item Type checking is decidable.
\end{itemize}
These are very strong constraints, which interact with each other in subtle ways and place strong restrictions on the way mathematics is represented and carried out. Not every proof assistant adopts such a framework. In fact, most reject the third, allowing classical reasoning that is ubiquitous in contemporary mathematics. Similarly, the propositions-as-types correspondence is usually linked to constructive theories, though there is no reason that it cannot be adopted in classical frameworks as well.

Although the mechanisms for type inference described in this article scale reasonably well, their use in real mathematical settings can be complex and delicate. Moreover, when an expression fails to typecheck, error messages from the system are often uninformative, and it can be frustrating and difficult to diagnose the problem. There are, moreover, rigid limitations to dependent type theory that stem from the commitment to keep type checking decidable. This is so because type checking algorithms are constrained to focus on syntactic structure, without incorporating background knowledge. For example, if \form{list A n} denotes the type of vectors of elements of \form{A} of length \form{n}, and we have \form{t\ :\ list A (0 + n)}, then, in Coq, \form{t} also typechecks as an element of \form{list A n}. In other words, Coq recognizes these two types as being the same. But entering \form{t\ :\ list A (n + 0)} yields a type error; Coq refuses to recognize that \form{list A (n + 0)} is the same as \form{list A n}. What is going on is that addition in Coq is defined by recursion on the first argument, so that the the term \form{0 + n} reduces to \form{n} under the computational interpretation. But the fact that \form{n + 0} is equal to \form{0} is a \emph{mathematical} fact, and there is no general way to incorporate arbitrary mathematical information in type checking while maintaining decidability.

Still, some have explored ways of making type judgments more flexible while maintaining decidability \cite{altenkirch:et:al:07,blanqui:et:al:99,strub:10}. An alternative is to give up the decidability of type checking, and accept the fact that some type judgments will require proof from the user. This is the path chosen by NuPrl \cite{constable:86} and PVS \cite{shankar:owre:00}. Yet another alternative is to jettison type theory altogether, and move to an axiomatic system like set theory, which offers maximum flexibility while relinquishing all the benefits of types; and then try to recapture some of those benefits by adding an extra layer of automation to register and manage domain information outside the axiomatic theory. Such ``soft typing'' mechanisms can be found, for example, in Mizar \cite{grabowski:et:al:10}.

This article has focused on the modeling of mathematical language from the point of view of contemporary interactive theorem provers. Others \cite{cramer:et:al:11,ganesalingam:09} have come at the problem from the perspective of natural language processing. In the long run, it seems likely that the various approaches will converge.

Inferring domain information is essential to modeling mathematical language and reasoning. Gonthier's work on the Feit-Thompson theorem shows that it is possible to model full-blown algebraic reasoning in an interactive proof systems. But other approaches should also be explored, and continued experimentation and innovation is needed to develop better support for verifying ordinary mathematical proofs.


\end{document}